\documentclass[10pt]{amsart}

\usepackage{amsmath,amssymb,amsfonts}

\def\dis
{\displaystyle}

\def\qed
{\hfill $\square$}

\def\R{{\mathbb R}}

\def\Tend#1#2{\mathop{\longrightarrow}\limits_{#1\rightarrow#2}}

\def\d{{\partial}}

\newtheorem{lem}{Lemma}[section]
\newtheorem{cor}[lem]{Corollary}
\newtheorem{prop}[lem]{Proposition}
\newtheorem{theo}[lem]{Theorem}

\theoremstyle{remark}
\newtheorem{remark}[lem]{Remark}
\newtheorem{ex}[lem]{Example}

\def\dem
{\noindent {\it Proof.} }

\numberwithin{equation}{section}

\begin{document}
\title[Critical NLS with and
without harmonic potential]{Critical nonlinear Schr\"odinger equations
with and without harmonic potential}
\author[R. Carles]{R\'emi Carles }                     

\address{Math\'ematiques Appliqu\'ees de Bordeaux et UMR 5466 CNRS,
Universit\'e Bordeaux 1, 351 cours de la Lib\'eration, 33~405 Talence
cedex, France}
\email{carles@math.u-bordeaux.fr}
\begin{abstract}
We use a change of variables that turns the critical nonlinear
Schr\"odinger equation into the critical nonlinear
Schr\"odinger equation with isotropic harmonic potential, in 
any space dimension. This change
of variables is isometric on $L^2$, and bijective on some time
intervals. Using the known results for the critical nonlinear
Schr\"odinger equation, this provides information for the properties of
Bose-Einstein condensate in space dimension one and two. We discuss in
particular the wave collapse phenomenon.  
\end{abstract}
\maketitle
\section{Introduction}
\label{sec:intro}

Bose-Einstein condensation is usually modeled by a nonlinear
Schr\"odinger equation with harmonic potential (see e.g. \cite{CCT}),
$$ i\hbar\d_t \psi^\hbar +\frac{\hbar^2}{2m}\Delta \psi^\hbar 
=\frac{m}{2}\omega^2x^2 \psi^\hbar +\frac{4\pi\hbar^2 a}{m}
|\psi^\hbar|^2 \psi^\hbar,\ \ \ (t,x)\in \R\times\R^n,$$ 
where $\omega >0$ and $a$ is the scattering length, whose sign differs
according to the chemical element considered. For instance, it is negative for
$\,^7$Li atoms (\cite{BSTH}, \cite{BSH}), as well as for $\,^{85}$Rb, and
positive for $\,^{87}$Rb, $\,^{23}$Na and $\,^1$H. The harmonic
potential $x^2$ models a magnetic field whose role is to confine the
particles (this is one of the ingredients for Bose-Einstein
condensation, once the atoms have been cooled by a laser, see e.g. 
\cite{BEC}), 
and the nonlinear term takes the (main) interactions between
the particles into account.  
To simplify the
mathematical analysis, we assume from now on that $m=\hbar=1$, and we
denote $4\pi a$ by $\lambda \in \R$. Notice that in  \cite{XEDP} and
\cite{AHP}, we considered the semi-classical limit $\hbar \rightarrow
0$. 

In the above equation, the nonlinearity is cubic, regardless of the
space dimension $n\geq 1$. Other models are also considered. In \cite{KNSQ},
the authors propose a quintic nonlinearity in space dimension one, and
in \cite{Zhang}, the author suggests more generally the study of 
\begin{equation}\label{eq:general}
\left\{
\begin{split}
i\d_t u +\frac{1}{2}\Delta u & = \frac{\omega^2x^2}{2}u + \lambda
|u|^{4/n}u,\ \ \ (t,x)\in \R\times\R^n,\\ 
u_{\mid t=0}&=u_0.
\end{split}
\right.
\end{equation}
As noticed in \cite{Zhang}, the proposed nonlinearity is the usual
critical nonlinearity for the nonlinear Schr\"odinger equation with no
potential ($\omega =0$, see e.g. \cite{Caz}). When $n=1$, this
suggestion meets the model proposed in \cite{KNSQ}, and when $n=2$,
this is the usual cubic nonlinearity. The results in \cite{Zhang}
enlighten a rather surprising analogy between the study of
(\ref{eq:general}) and that of 
\begin{equation}\label{eq:nls}
\left\{
\begin{split}
i\d_t v +\frac{1}{2}\Delta v & =   \lambda
|v|^{4/n}v,\ \ \ (t,x)\in \R\times\R^n,\\ 
v_{\mid t=0}&=u_0.
\end{split}
\right.
\end{equation}
Many results are known for (\ref{eq:nls}), we recall some of
them. In \cite{Weinstein}, the author proved that if 
$$u_0 \in \Sigma := H^1(\R^n)\cap \left\{\phi \in L^2(\R^n);\ |x|\phi
\in L^2(\R^n) \right\},$$ 
then there exists $T>0$ such that $v\in C(]-T,T[,\Sigma)$. If $\lambda
\geq 0$, then one can take $T=\infty$. When $\lambda <0$, one can take
$T=\infty$ when $\|u_0\|_{L^2}$ is sufficiently small. More precisely,
let $Q$ denote the ground state, which is the unique radial solution
of (see \cite{Strauss}, \cite{Kwong})
\begin{equation}\label{eq:Q}
\left\{
\begin{split}
-\frac{1}{2}\Delta Q + Q  =-\lambda|Q|^{4/n}Q,& \textrm{ in }\R^n, \\
Q >0,& \textrm{ in }\R^n.
\end{split}
\right.
\end{equation}
Weinstein proved that if $\|u_0\|_{L^2}< \|Q\|_{L^2}$, then one can take 
$T=\infty$. On the other hand, if $ \|u_0\|_{L^2}\geq \|Q\|_{L^2}$,
then the wave $v$ may collapse in finite time. Zhang proved that the same
holds for the solution $u$ of (\ref{eq:general}). We use a change
of variables that shows why this is so. 

Fix $\omega >0$. Let $v$ be a solution of (\ref{eq:nls}), for $|t|<T$, and
define, for $|t|<\arctan(\omega T)/\omega$, 
\begin{equation}\label{eq:chgt1}
u(t,x)= \frac{1}{(\cos \omega t)^{n/2}}e^{-i\frac{\omega}{2}x^2 \tan \omega
t} v\left(\frac{\tan \omega t}{\omega},\frac{x}{\cos \omega t} \right).
\end{equation}
Then $u$ solves (\ref{eq:general}). This was first noticed in
\cite{Niederer} for the linear case ($\lambda =0$), and in
\cite{Rybin} for the nonlinear case with critical nonlinearity. 
Reciprocally, if $u$ solves
(\ref{eq:general}), then $v$, defined by
\begin{equation}\label{eq:chgt2}
v(t,x)=\frac{1}{\left(1+ (\omega t)^2 \right)^{n/4}}e^{i\frac{\omega^2
t}{1+(\omega t)^2}\frac{x^2}{2} }u\left( \frac{\arctan \omega t}{\omega},
\frac{x}{\sqrt{ 1 + (\omega t)^2}}\right),
\end{equation}
solves (\ref{eq:nls}). 
The transforms (\ref{eq:chgt1}) and (\ref{eq:chgt2}) do
not alter the initial data $u_0$, and are isometric on
$L^2(\R^n)$. Therefore,  it
is not surprising that global existence is obtained when
$\|u_0\|_{L^2}< \|Q\|_{L^2}$ for both cases, provided that it is known
for one of them. In Sect.~\ref{sec:known}, we recall the known results
for Eq.~(\ref{eq:nls}), and analyze the relations between
(\ref{eq:general}) and (\ref{eq:nls}) in Sect.~\ref{sec:rel}. Finally,
we investigate the consequences of the results known for
(\ref{eq:nls}) as far as Bose-Einstein equations are concerned, in
Sect.~\ref{sec:trad}.

\section{Known results for the critical nonlinear Schr\"odinger
equation} 
\label{sec:known}

Start with the initial value problem. It is now classical that if
$u_0\in H^1(\R^n)$, then there exists a solution of the initial value
problem (\ref{eq:nls}), continuous on some (possibly small) time
interval, with values in $H^1(\R^n)$ (see e.g. \cite{Caz}). The same
holds if $u_0\in \Sigma$  (which is
the natural space to study (\ref{eq:general}), since $\Sigma = D\left(
\sqrt{-\Delta +x^2}\right)$). Moreover, the mass and
energy associated to 
the equation are conserved. Because we consider the nonlinear
Schr\"odinger equation with \emph{critical} power, the
pseudo-conformal conservation law is in our case an exact conservation
law. 

\begin{prop}\label{prop:ivpnls}
For every $u_0\in H^1(\R^n)$ (resp. $u_0\in \Sigma$), there exist
$T_*(u_0)$, $T^*(u_0)>0$ and there exists a unique, maximal solution
\begin{equation*}
\begin{split}
v\in & C(]-T_*(u_0),T^*(u_0)[,H^1(\R^n))\cap
C^1(]-T_*(u_0),T^*(u_0)[,H^{-1}(\R^n))\\
(\textrm{resp. }v\in &
C(]-T_*(u_0),T^*(u_0)[,\Sigma)\cap  
C^1(]-T_*(u_0),T^*(u_0)[,H^{-1}(\R^n))) 
\end{split} 
\end{equation*} 
of problem 
(\ref{eq:nls}). The solution $v$ is maximal in the sense that if
$T^*(u_0)<\infty$, then $\|v(t)\|_{H^1} \rightarrow \infty$ as
$t\uparrow T^*(u_0)$, and if $T_*(u_0)<\infty$, then $\|v(t)\|_{H^1}
\rightarrow \infty$ as 
$t\downarrow -T_*(u_0)$. In addition, we have the following three
conservation laws for $t\in ]-T_*(u_0),T^*(u_0)[$.\\
1. Conservation of mass: $\|v(t)\|_{L^2}=\|u_0\|_{L^2}$. \\
2. Conservation of energy: 
$$E_1(t):= \frac{1}{2}\|\nabla_x
v(t)\|_{L^2}^2 + \frac{\lambda}{1+2/n}\| v(t)\|_{L^{2+4/n
}}^{2+4/n}\equiv E_1(0).$$
3. Pseudo-conformal conservation law:
$$E_2(t) := \frac{1}{2}\|(x+it\nabla_x)
v\|_{L^2}^2 + \frac{\lambda t^2}{1+2/n}\|
v(t)\|_{L^{2+4/n}}^{2+4/n}\equiv E_2(0) .$$
\end{prop}
\begin{remark}
The lower regularity $u_0\in L^2$ could be considered as well, using
the results of Cazenave and Weissler \cite{CW}. 
\end{remark}

In some cases, it is known that we have $T_*(u_0)=T^*(u_0)=\infty$ (see
e.g. \cite{Caz}). 
\begin{prop}
\label{prop:global}
Let $u_0 \in H^1(\R^n)$ (resp. $u_0\in \Sigma$), and let $v$ be the maximal
solution of (\ref{eq:nls}). Then $v$ is defined globally in time, that
is $T_*(u_0)=T^*(u_0)=\infty$, in either of the following cases. 
\begin{itemize}
\item If the nonlinearity is repulsive, $\lambda >0$. 
\item If the nonlinearity is attractive, $\lambda <0$, and the mass of
$u_0$ is sub-critical in the sense that $\|u_0\|_{L^2}<\|Q\|_{L^2}$,
where $Q$ is the ground state defined by (\ref{eq:Q}). 
\end{itemize}
\end{prop}

On the other hand, we have some sufficient conditions for which it is
known that the solution blows up in finite time. We restrict to the
case $T^*(u_0)<\infty$, which corresponds to a wave collapse in the
future. 

\begin{prop}[\cite{Weinstein}, Th.~4.2] 
\label{prop:blow}Recall that $E_1$ denotes
the energy associated to (\ref{eq:nls}), that it is constant. Let either

(i) $E_1<0$,

(ii) $E_1=0$ and $\operatorname{Im}\int \overline{u_0}x.\nabla u_0 dx
<0$,\\
or

(iii) $E_1>0$ and $\operatorname{Im}\int \overline{u_0}x.\nabla u_0 dx
\leq -2 \sqrt {E_1}\|x u_0\|_{L^2}$.\\
Then there exists $0<T<\infty$ such that 
$$\lim_{t\rightarrow T}\|\nabla_x v(t)\|_{L^2}=\infty.$$
\end{prop}

We now assume that $\lambda<0$: global existence is not guaranteed. 
When blow up occurs (which could be the case under other conditions
than those stated in Prop.~\ref{prop:blow}),
Merle analyzed very precisely its mechanism, when the mass is
critical. From Prop.~\ref{prop:global}, global existence is ensured
when $\|u_0\|_{L^2}<\|Q\|_{L^2}$. Weinstein \cite{Weinstein} proved
that if $\|u_0\|_{L^2}=\|Q\|_{L^2}$, then wave collapse may occur, at
least for some particular initial data. Merle proved that up to the
invariants of (\ref{eq:nls}), the blowing up solutions enlightened by
Weinstein are the only ones. 
\begin{theo}[\cite{MerleDuke}, Th.~1]\label{th:Merle}
 Let $\lambda <0$, $u_0\in
H^1(\R^n)$, and assume that the solution $v$ of (\ref{eq:nls}) blows
up in finite time 
$T>0$. Moreover, assume that $\|u_0\|_{L^2}=\|Q\|_{L^2}$, where $Q$ is
defined by (\ref{eq:Q}). Then there
exist $\theta \in \R$, $\delta >0$, $x_0,x_1 \in \R^n$ such that
$$u_0(x)=\left(\frac{\delta}{T}\right)^{n/2}e^{i\theta - i|x-x_1|^2/2T +
i\delta^2/T}Q\left( \delta \left(\frac{x-x_1}{T}-x_0
\right)\right),$$
and for $t<T$, 
$$v(t,x)=\left(\frac{\delta}{T-t}\right)^{n/2}e^{i\theta -
i|x-x_1|^2/2(T-t) + 
i\delta^2/(T-t)}Q\left( \delta \left(\frac{x-x_1}{T-t}-x_0
\right)\right).$$
\end{theo}
\begin{remark}\label{rq:disp}
In addition, Merle proved that when the mass is critical, only
three causes can prevent the global definition of $v$ with optimal
dispersion of the $L^{2+4/n}$-norm of $v$ (\cite{MerleDuke},
Cor.~1.2).
\begin{itemize}
\item The initial data of Th.~\ref{th:Merle}, that cause blow up at some
positive time.
\item Their conjugates, that cause blow up at some
negative time.
\item The solitary waves, caused by
$$u_0(x) = \delta^{n/2}e^{i\theta}Q\left( \delta  (x-x_0
)\right).$$
\end{itemize}
\end{remark}
We conclude this section by recalling another consequence of
Th.~\ref{th:Merle}. 
\begin{cor}[\cite{MerleDuke}, Cor.~1.1]
\label{cor:Merle} Let $\lambda<0$. 
The solutions of the Cauchy problem
\begin{equation}
\left\{
\begin{split}
i\d_t v +\frac{1}{2}\Delta v & =  \lambda
|v|^{4/n}v,\ \ \ \textrm{ for }t>0,\\ 
|v(0,x)|^2&=\|Q\|_{L^2}^2\delta_{x=0},
\end{split}
\right.
\end{equation}
are exactly, for $\theta \in \R$, $\delta >0$, $x_0\in \R^n$, 
\begin{equation*}
v(t,x)= \left(\frac{\delta}{t} \right)^{n/2}e^{i\theta + i\frac{\delta^2}{t}
-i\frac{x^2}{2t}} Q \left( \frac{\delta}{t}x -x_0\right). 
\end{equation*}
\end{cor}

\section{Transformation and relation between the two equations}
\label{sec:rel}
Let $u_0\in \Sigma$. From Prop.~\ref{prop:ivpnls}, the problem
(\ref{eq:nls}) has a unique solution $v \in C(]-T,T[,\Sigma)$ for some
positive $T$. For $|t|<\arctan(\omega T)/\omega$, let $u$ be defined
by (\ref{eq:chgt1}), that is  
\begin{equation*}
u(t,x)= \frac{1}{(\cos \omega t)^{n/2}}e^{-i\frac{\omega}{2}x^2 \tan \omega
t} v\left(\frac{\tan \omega t}{\omega},\frac{x}{\cos \omega t} \right).
\end{equation*}
Then $u \in C(]-\arctan(\omega T)/\omega,\arctan(\omega
T)/\omega[,\Sigma)$, and $u$ solves (\ref{eq:general}). Conversely, if
$u \in C(]-\tau,\tau[,\Sigma)$ solves (\ref{eq:general}) with
$0<\tau<\pi/2$, then $v$ 
defined by (\ref{eq:chgt2}), that is
\begin{equation*}
v(t,x)=\frac{1}{\left(1+ (\omega t)^2 \right)^{n/4}}e^{i\frac{\omega^2
t}{1+(\omega t)^2}\frac{x^2}{2} }u\left( \frac{\arctan \omega t}{\omega},
\frac{x}{\sqrt{ 1 + (\omega t)^2}}\right),
\end{equation*}
is such that $v \in C(]-\tan (\omega \tau)/\omega,\tan (\omega
\tau)/\omega[,\Sigma)$. Transposing similarly the results of
Prop.~\ref{prop:ivpnls} yields the following corollary,

\begin{cor}\label{cor:ivp}
Let $u_0\in \Sigma$. Then  
there exist
$\tau_*(u_0)$, $\tau^*(u_0)>0$ and there exists a unique, maximal solution
\begin{equation*} 
u\in 
C(]-\tau_*(u_0),\tau^*(u_0)[,\Sigma)\cap  
C^1(]-\tau_*(u_0),\tau^*(u_0)[,H^{-1}(\R^n))
\end{equation*} 
of problem 
(\ref{eq:general}). It is maximal in the sense that is
$\tau^*(u_0)<\infty$, then $\|u(t)\|_{H^1} \rightarrow \infty$ as
$t\uparrow \tau^*(u_0)$, and if $\tau_*(u_0)<\infty$, then $\|u(t)\|_{H^1}
\rightarrow \infty$ as 
$t\downarrow -\tau_*(u_0)$. In addition, the following three
quantities are constant for $t\in ]-\tau_*(u_0),\tau^*(u_0)[$.\\
1. Conservation of mass: $\|u(t)\|_{L^2}=\|u_0\|_{L^2}$. \\
2. First part of the energy: 
\begin{equation}\label{eq:E_1}
E_1(u)= \frac{1}{2}\left\| \omega x \sin \omega t u(t)-i \cos \omega t
\nabla_x u(t)\right\|^2_{L^2} +\frac{n\lambda}{n+2}\cos^2 \omega t
\|u(t)\|_{L^{2+4/n}}^{2+4/n}.
\end{equation}
3. Second part of the energy: 
\begin{equation}\label{eq:E_2}
E_2(u)= \frac{1}{2}\left\| \omega x \cos \omega t u(t)+i \sin \omega t
\nabla_x u(t)\right\|^2_{L^2} +\frac{n\lambda}{n+2}\sin^2 \omega t
\|u(t)\|_{L^{2+4/n}}^{2+4/n}.
\end{equation}
\end{cor}
\begin{remark}
The existence part of this result was proved in \cite{Oh}, and
revisited in \cite{AHP}. In the special case of a critical power
nonlinearity, the transform (\ref{eq:chgt1}) shows that no new proof
is needed when Prop.~\ref{prop:ivpnls} is known. 
\end{remark}
\dem The only point that we have to prove is the blow up
case. Assume for instance that $\tau^*(u_0)$ is finite. Up to a time
translation, we can suppose that $0<\tau^*(u_0)<\pi/2\omega$. Then
$v$, defined by (\ref{eq:chgt2}), solves (\ref{eq:nls}). Let $\tilde
v$ be the maximal solution of (\ref{eq:nls}) given by
Prop.~\ref{prop:ivpnls}. If it were globally defined, then $v$ would
also be globally defined; the transform  (\ref{eq:chgt1}) would then
make it possible to define $u$ up to some time $\tau$ such that
$\tau^*(u_0)<\tau<\pi/2\omega$, which contradicts the maximality of
$\tau^*(u_0)$. Therefore, there exists $T^*<\infty$ such that 
$$\|\nabla_x \tilde v(t)\|_{L^2} \Tend t {T^*} \infty.$$
We prove that $u$ blows up at least before time $\arctan(\omega
T^*)/\omega$. For  $0<t<\arctan(\omega T^*)/\omega$, 
\begin{equation*}
\begin{split}
\|\nabla_x u(t)\|_{L^2} &= \left\| -i\omega x\sin (\omega
t)v\left(\frac{\tan \omega t}{\omega},.\right) +  \nabla_x
v\left(\frac{\tan \omega t}{\omega},.\right)\right\|_{L^2}\\
&\geq \left\|\nabla_x v\left(\frac{\tan \omega
t}{\omega},.\right)\right\|_{L^2} -\left\|\omega x\sin (\omega
t) v\left(\frac{\tan \omega
t}{\omega},.\right)\right\|_{L^2}. 
\end{split}
\end{equation*}
Now we know that for $t<T^*$ (see for instance \cite{Weinstein} and
references therein),  
$$\frac{d^2}{dt^2}\|x v(t,x)\|_{L^2} = 4 E_1,$$
where $E_1$ denotes the energy of $v$, which is constant. Then letting
$t$ go to 
$$\tau^*=\frac{\arctan\omega T^*}{\omega}$$
 yields the blow up part of the corollary. \qed

\begin{remark}
Both conservation laws (\ref{eq:E_1}) and (\ref{eq:E_2}) were derived
in \cite{AHP}, in the case of a more general nonlinearity, not
necessarily critical (the second
terms of $E_1(u)$ and $E_2(u)$ have to be adapted according to the
power considered); in general, $E_1(u)$ and $E_2(u)$ do depend on
time, they are constant only in the case of a critical power. On the
other hand, the sum of $E_1(u)$ and $E_2(u)$ is always constant, and
corresponds to the usual energy associated to (\ref{eq:general}), 
$$E(u)= \frac{1}{2}\left\| \nabla_x u(t)\right\|^2_{L^2} +
\frac{1}{2}\left\| \omega x u(t)\right\|^2_{L^2}
+\frac{n\lambda}{n+2}
\|u(t)\|_{L^{2+4/n}}^{2+4/n}.$$
Notice that the energy for $v$ is always conserved as well (it
reflects the Hamiltonian structure), while the pseudo-conformal
conservation law is in general an evolution law, which is an exact
conservation law only in the critical case (and the free case $\lambda
=0$). 
\end{remark}

\section{Transposition of the results for Bose Einstein equations}
\label{sec:trad}
In this last section, we investigate some consequences of the
transform (\ref{eq:chgt1}) and its companion (\ref{eq:chgt2}) for Bose
Einstein equations. We consider in particular the questions of global
existence and wave collapse phenomenon.

Transposing Prop.~\ref{prop:global} yields the following corollary,  
\begin{cor}\label{cor:global} 
Let $u_0 \in \Sigma$, and $u$ the maximal 
solution of (\ref{eq:general}). Then $u$ is defined globally in time,
that 
is $\tau_*(u_0)=\tau^*(u_0)=\infty$, in either of the following cases.  
\begin{itemize} 
\item If the nonlinearity is repulsive, $\lambda >0$.  
\item If the nonlinearity is attractive, $\lambda <0$, and the mass of 
$u_0$ is sub-critical in the sense that $\|u_0\|_{L^2}<\|Q\|_{L^2}$, 
where $Q$ is the ground state defined by (\ref{eq:Q}).  
\end{itemize} 
\end{cor} 
\begin{remark} 
As mentioned in the introduction, this result was proved by Zhang 
\cite{Zhang}. We believe that our approach provides a good explanation 
for this result. 
\end{remark} 
\dem The repulsive case is straightforward, since the three 
conservations stated in Cor.~\ref{cor:ivp} provide \emph{a priori} 
bounds on the $\Sigma$-norm of $u$. When the nonlinearity is 
attractive and the mass of  
$u_0$ is sub-critical, we use Prop.~\ref{prop:global}. From the 
results of Weinstein \cite{Weinstein}, the solution of  
(\ref{eq:nls}) is defined globally when the mass of $u_0$ is 
sub-critical, $\|u_0\|_{L^2}< \|Q\|_{L^2}$. Then $u\in 
C([0,\pi/2\omega[,\Sigma)$, from the transform 
(\ref{eq:chgt1}). Recall that the $L^2$-norm of $u$ is  
constant on $[0,\pi/2\omega[$. Considering 
$t_0=\pi/4\omega$ as a new time origin and repeating this procedure 
shows that $u\in 
C([0,3\pi/4\omega[,\Sigma)$. Using this argument indefinitely  
shows that $u\in 
C([0,\infty[,\Sigma)$, and similarly, $u\in 
C(]-\infty,\infty[,\Sigma)$.~\qed

Like for the case of (\ref{eq:nls}), we can state sufficient
conditions where blow up occurs. This is done by transposing
Prop.~\ref{prop:blow} with (\ref{eq:chgt1}). 
\begin{cor}Recall that $E_1$ is defined by (\ref{eq:E_1}) and is constant as
long as $u$ belongs to $\Sigma$.  
Let either

(i) $E_1<0$,

(ii) $E_1=0$ and $\operatorname{Im}\int \overline{u_0}x.\nabla u_0 dx
<0$,\\
or

(iii) $E_1>0$ and $\operatorname{Im}\int \overline{u_0}x.\nabla u_0 dx
\leq -2 \sqrt {E_1}\|x u_0\|_{L^2}$.\\
Then there exists $0<\tau<\pi/2\omega $ such that 
$$\lim_{t\rightarrow \tau}\|\nabla_x u(t)\|_{L^2}=\infty.$$
\end{cor}
\begin{remark}
The above criteria yield wave collapse at time $\tau <\pi/2\omega
$. It is sensible to expect this phenomenon to occur possibly at time $\tau
=\pi/2\omega $, which corresponds to a focus for the free equation
(\ref{eq:general}) with $\lambda=0$. This geometric aspect is hidden
in the case of (\ref{eq:nls}), since it corresponds to infinite
times. We will consider this point more precisely later, in the case
of a critical mass 
($\|u_0\|_{L^2}=\|Q\|_{L^2}$). As noticed in \cite{Zhang} and
\cite{AHP}, wave collapse  for $u$ always occurs at time $\tau \leq
\pi/2\omega$  when $E_1=0$.  Thus we could say that the
compactification 
of time in the transformation (\ref{eq:chgt1}) leads
to new blowing up solutions. 
\end{remark}

On the other hand, if $\lambda<0$ and $\|u_0\|_{L^2}= \|Q\|_{L^2}$, then
wave collapse occurs. This can be analyzed very precisely thanks to
the results of Merle.  As a corollary of Th.~\ref{th:Merle}, 
with  the change of
variable $\delta ' =\delta \cos \omega \tau$, we have the following, 
\begin{cor}\label{cor:blow1}
Let $\lambda <0$, $u_0\in
H^1(\R^n)$, and assume that the solution $u$ of (\ref{eq:general}) blows
up in finite time 
$0<\tau<\pi/2\omega$. Moreover, assume that
$\|u_0\|_{L^2}=\|Q\|_{L^2}$. Then there
exist $\theta \in \R$, $\delta >0$, $x_0,x_1 \in \R^n$ such that
\begin{equation}\label{eq:data1}
%\begin{split}
u_0(x)=\left(\frac{\omega\delta}{\sin \omega \tau}\right)^{n/2}
e^{i\theta + 
i\frac{2\omega\delta^2}{ \sin 2\omega\tau}-
i\omega\frac{|x-x_1|^2}{2}\cot \omega\tau}
%& \times 
Q\left( \omega\delta \left(\frac{x-x_1}{\sin
\omega \tau}-\frac{x_0}{\omega}
\right)\right),
%\end{split}
\end{equation}
and for $0<t<\tau$, 
\begin{equation*}
\begin{split}
u(t,x)= &e^{i\theta + 
i\delta^2 \frac{\omega\cos \omega t}{\cos\omega\tau \sin
\omega (\tau -t)} -i \frac{\omega}{2}\left|\frac{x}{\cos \omega t}-x_1
 \right|^2\frac{\cos \omega t \cos\omega\tau }{\sin
\omega (\tau -t)}-i\omega \frac{x^2}{2}\tan \omega t}\times \\
& \times 
\left(\frac{\omega\delta
}{\sin\omega(\tau-t)}\right)^{n/2}Q\left( \omega\delta \left(\frac{x -x_1
\cos\omega 
t}{\sin \omega(\tau-t)}-\frac{x_0}{ \omega}
\right)\right).
\end{split}
\end{equation*}
\end{cor}
\begin{ex}
Blow up at time $\tau =\pi/4\omega$, with critical mass
$\|u_0\|_{L^2}=\|Q\|_{L^2}$, is caused by initial data of the form
\begin{equation*}
u_0(x)=\delta^{n/2}
e^{i\theta - 
i\omega\frac{|x-x_1|^2}{2}
} Q\left( \delta  \left(x-x_1-x_0
\right)\right). 
\end{equation*}
Recall that the quadratic oscillations always cause a focus for
(\ref{eq:nls}) (see e.g. \cite{Indiana}). On the other hand, the
geometry of the harmonic potential creates a focus at time
$\pi/2\omega$. Therefore we can say that in the above case, both
phenomena cumulate,  to anticipate the ``usual'' blow up. 
\end{ex}

\begin{ex}
Using a time translation shows that
blow up at time $\tau =3\pi/4\omega$, with critical mass
$\|u_0\|_{L^2}=\|Q\|_{L^2}$, is caused by initial data of the form
\begin{equation*}
u_0(x)=\delta^{n/2}
e^{i\theta +
i\omega\frac{|x-x_1|^2}{2}
} Q\left( \delta  \left(x-x_1-x_0
\right)\right). 
\end{equation*}
In that case, the oscillation $e^{i\omega\frac{x^2}{2}}$ tends to delay the
concentration (it is the outgoing oscillation once the focus has been
crossed in \cite{Indiana}), but the geometry of the harmonic potential
counterbalances this phenomenon and eventually causes wave collapse. 
\end{ex}
\begin{remark}\label{rq:pi}
We could of course state the analogue of this result for blow up in
the past, $-\pi/2\omega < \tau <0$. Since the transform
(\ref{eq:chgt1}) is ``almost'' $\pi$-periodic, and using the fact that
the ground state $Q$ is spherically symmetric, we can deduce in
particular that if $u(t)$ does not blow up for $t\in [0,\pi/\omega[$,
then it will never blow up in the future, and has not collapsed in the
past.
\end{remark}

From Remark~\ref{rq:disp}, three causes can prevent the global
definition of solutions of (\ref{eq:nls}) with optimal 
dispersion of the $L^{2+4/n}$-norm of $v$, when the mass is critical. 
We have not analyzed the last possibility yet. In the case of solitary
waves, we have
$$v(t,x)= \delta^{n/2}e^{i\theta}Q\left( \delta  (x-x_0
)\right)e^{i\delta^2t}.$$
The transformation (\ref{eq:chgt1}) then yields
\begin{equation}\label{eq:data2}
u(t,x)=\left(\frac{\delta}{\cos \omega t}\right)^{n/2}e^{i\theta
+i\omega \tan \omega t \left(\frac{\delta^2}{\omega^2}
-\frac{|x|^2}{2}\right)
 }Q\left(\delta \left( \frac{x}{\cos \omega t}-x_0\right)\right).
\end{equation}
From \cite{MerleDuke}, Cor.~1.2, all the
initial data $u_0$ with 
$\|u_0\|_{L^2} = \|Q\|_{L^2}$ different from (\ref{eq:data1}) and
(\ref{eq:data2}) yield a solution $v$ globally defined, with
$\|v(t)\|_{L^{2+4/n}}^{2+4/n}=O(t^{-2})$. Back to $u$, thanks to the
transformation (\ref{eq:chgt1}), this provides a uniform estimate for
$\|u(t)\|_{L^{2+4/n}}^{2+4/n}$, when $t\in [0,\pi/2\omega[$. From the
conservation of the energy $E(u)$ (the usual Hamiltonian), along with
the conservation of mass, this yields
an {\it a priori} estimate for the $\Sigma$-norm of
$u(t,.)$. Therefore, $u$ does not blow up at time $\pi/2\omega$
(otherwise, its $H^1$-norm would not be bounded near $\pi/2\omega$,
see e.g. \cite{Caz}, Th.~4.2.8). From Prop.~\ref{prop:ivpnls}, there
exists some positive $\alpha$ such that $u$ is defined for 
$t\in [0, \pi/2\omega +\alpha[$. One can use similar arguments for
negative times, and using a time translation, we can deduce a similar
description for $t\in [0, \pi/\omega[$. Finally, if the solution has
not blown up when reaching $t=\pi/\omega$, then it will never blow up,
as mentioned in Remark~\ref{rq:pi}. To summarize, we have the following,

\begin{cor}\label{cor:blow2}
Let $\lambda <0$, and $u_0\in \Sigma$ be such that
$\|u_0\|_{L^2}=\|Q\|_{L^2}$. Assume in addition that 
\begin{equation*}
u_0(x) \not\equiv \delta^{n/2}
e^{i\theta -i\omega \cot \omega\tau
\frac{|x-x_1 |^2}{2}}
 Q\left( \delta  \left(x-x_1-x_0\right)\right)
\end{equation*}
for $(\delta,\theta,\tau,x_0,x_1)\in \R^*_+\times \R \times \dis
\left]0,\frac{\pi}{2\omega}  \right[\times
\R^n\times\R^n$. Then $u(t)$ is defined for $t\in [0,\pi/2\omega[$. If
in addition, 
\begin{equation*}
u_0(x)\not\equiv \delta^{n/2}e^{i\theta}Q\left(\delta (
x-x_0)\right) 
\end{equation*}
for $(\delta,\theta,x_0)\in \R^*_+\times \R \times 
\R^n$, then $u(t)$ is defined for $t\in [0,\pi/2\omega+\alpha[$ for
some positive $\alpha$. Finally, if moreover 
\begin{equation*}
u_0(x) \not\equiv \delta^{n/2}
e^{i\theta -i\omega \cot \omega\tau
\frac{|x-x_1 |^2}{2}}
 Q\left( \delta  \left(x-x_1-x_0\right)\right)
\end{equation*}
for $(\delta,\theta,\tau,x_0,x_1)\in \R^*_+\times \R \times \dis
\left]0,\frac{\pi}{\omega}  \right[\times
\R^n\times\R^n$, then $u(t)$ is defined for $t\in ]-\infty,\infty[$.  
\end{cor}

Adapting Cor.~\ref{cor:Merle} yields the next
corollary.
\begin{cor}\label{cor:dirac}
Let $\lambda <0$. The solutions of the Cauchy problem
\begin{equation}
\left\{
\begin{split}
i\d_t u +\frac{1}{2}\Delta u & = \frac{\omega^2x^2}{2}u + \lambda
|u|^{4/n}u,\ \ \ \textrm{ for }t>0,\\ 
|u(0,x)|^2&=\|Q\|_{L^2}^2\delta_{x=0},
\end{split}
\right.
\end{equation}
are exactly, for $\theta \in \R$, $\delta >0$, $x_0\in \R^n$, $|\tau|
<\pi/2\omega$,  
\begin{equation*}
\begin{split}
u(t,x)= \left(\frac{\omega\delta}{\sin \omega t}\right)^{n/2}& e^{i\theta
+i\omega\delta^2\frac{\cos \omega (\tau -t)}{\cos \omega \tau
\sin \omega t} -i\omega\frac{x^2}{2}\left(\frac{\cos \omega \tau}{\cos
\omega (\tau -t)\sin \omega t}+ \tan \omega (\tau -t)\right)}\times \\
& \times Q\left(\omega\delta \left( \frac{x}{\sin \omega
t}-\frac{x_0}{\omega}\right)\right),
\end{split}
\end{equation*}
and 
\begin{equation*}
u(t,x)=\left(\frac{\omega\delta}{\sin \omega t}\right)^{n/2}e^{i\theta
+i\omega \cot \omega t \left(\frac{\delta^2}{\omega^2}
-\frac{|x|^2}{2}\right)
 }Q\left(\omega\delta \left( \frac{x}{\sin \omega
t}-\frac{x_0}{\omega}
\right)\right).
\end{equation*}
\end{cor}

\begin{remark}
When letting $\omega$ go to zero in Cor.~\ref{cor:blow1} and
Cor.~\ref{cor:dirac}, we retrieve the initial results of Merle. 
Letting $\tau$ go to $\pi/2\omega$ in the first part of
Cor.~\ref{cor:blow2} yields the second part of Cor.~\ref{cor:blow2},
but the same does not hold in Cor.~\ref{cor:dirac}. The two kinds of
solutions have a different structure (it is so in the case with no
potential). 
\end{remark}

\begin{remark}Continuation after blow-up time. In \cite{MerleCPAM},
Merle considers the possible continuations of the solution after the
breaking time. With (\ref{eq:chgt1}), we could adapt this theory to
the case of (\ref{eq:general}). However, it seems very likely that
(\ref{eq:general}) does not remain a good model for Bose-Einstein
condensation after the wave collapse.
\end{remark}

\begin{remark}
In \cite{MerleCMP} (see also \cite{MerleConf} for other results), it
is proved that one can fix $k$ points in $\R^n$ and 
construct a solution of (\ref{eq:nls}) that blows up exactly at these
points. This could be done for (\ref{eq:general}) as well. 
\end{remark}

\bibliographystyle{plain}
\bibliography{carles}

\end{document}